\documentclass{radioeng}
\usepackage[cp1250]{inputenc} 
\usepackage{graphicx} 
\usepackage{color}
\usepackage{cite}
\usepackage{listings}  
\usepackage{algorithmic}
\newcommand{\area}{
} 
\begin{document}
\setcounter{firstpage}{1}
\rengHeader{}{}{
~}{ \dots}
\rengTitle{Two Power Allocation and Beamforming Strategies for Active IRS-aided Wireless Network via Machine Learning}
\rengNames{Qiankun Cheng $^\mathit{1}$, Jiatong Bai $^\mathit{1}$, Baihua Shi $^\mathit{1,2}$, Wei Gao $^\mathit{1}$, Feng Shu $^\mathit{1,2}$}
\rengAffil{$^1$ School of Information and Communication Engineering, Hainan University, Haikou 570228, China \\
$^2$ School of Electronic and Optical Engineering, Nanjing University of Science and Technology, Nanjing 210094, China 
}
\rengMail{cqk1129@hainanu.edu.cn, 18419229733@163.com, shibh56h@foxmail.com, gaowei@hainanu.edu.cn, shufeng0101@163.com}
\begin{multicols}{2}

\begin{rengAbstract}
 This paper models an active intelligent reflecting surface (IRS) -assisted wireless communication network, which has the ability to adjust power between BS and IRS. We aim to maximize the signal-to-noise ratio of user by jointly designing power allocation (PA) factor, active IRS phase shift matrix, and beamforming vector of BS, subject to a total power constraint. To tackle this non-convex problem, we solve this problem by alternately optimizing these variables. Firstly, the PA factor is designed via polynomial regression method. Next, BS beamforming vector and IRS phase shift matrix are obtained by Dinkelbach's transform and successive convex approximation methods. To reduce the high computational complexity of the above proposed algorithm, we maximize achievable rate (AR) and use closed-form fractional programming method to transform the original problem into an equivalent form. Then, we address this problem by iteratively optimizing auxiliary variables, BS and IRS beamformings. Simulation results show that the proposed algorithms can effectively improve the AR performance compared to fixed PA strategies, aided by passive IRS, and without IRS.

\end{rengAbstract}
\rengKeywords{Active intelligent reflecting surface, achievable rate, power allocation, closed-form fractional programming}

\rengSection{Introduction}

With the rapid development of 5th generation (5G) communication technology and the emergence of a large number of new applications such as augmented reality and virtual reality, the demand for high quality and high speed wireless communication network is growing day by day \cite{g1,g2,g3}. However, high-rate wireless networks also face some new challenges, like high costs and high energy consumption. The realization of green wireless transmission has become the consensus of industry and academia \cite{g4,g5,g6}. As a low-cost, low-power reflector, intelligent reflecting surface (IRS) provides a new way for future green wireless communication \cite{g7}.

In an IRS-aided multiple-input single-output (MISO) system \cite{g8}, two optimization schemes were proposed to minimize the power consumption at base station (BS) by jointly optimizing the transmitting beamforming of BS and phase shift matrix of IRS, given the signal-to-noise (SNR) target of the receiving end user. In \cite{g9}, by using a deep reinforcement learning neural network, it was possible to simultaneously optimize transmit beamforming at BS and phase shift matrices at IRS to maximize ergodic sum rate in an IRS-aided multiuser downlink MISO system. In \cite{g10}, the authors introduced IRS into secure multiple-input multiple-output (MIMO) wireless powered communication networks. A secrecy rate maximization problem was investigated and alternating optimization methods were constructed based on mean-square error and dual subgradient techniques to solve this non-convex problem. Literatures \cite{g11} studied an IRS-assisted MIMO system, by iteratively optimizing the precoding beamformings in every BSs and phase shift beamforming in IRS, block coordinate descent and complex circle manifold methods are proposed to maximize the weighted sum rate. The deployment of IRS can significantly improve the performance of cell edge users compared to MIMO communication systems without IRS.

Although the power consumption of passive IRS mainly composed of passive reflective elements is significantly lower than that of active IRS, recent studies have shown that active IRS may have superiorities in some scenarios \cite{g12,g13,g14,g15}. The power gain achieved by passive IRS is limited in some cases due to the fact that double fading effect caused by signal transmission over the BS-to-IRS and IRS-to-user channels \cite{g16}. By using active IRS with power amplifiers, the impact of double fading can be reduced \cite{g17}. The authors in \cite{g18} compared the performance of active and passive IRS-assisted communication systems with the same total power budget, which proved that active IRS was superior to passive IRS when the power budget and the number of IRS elements are at a moderate level. The achievable rate (AR) maximization problem was studied in \cite{g19} in an active IRS-aided  single-input single-output system, closed-form maximum ratio reflecting and selective ratio reflecting methods were proposed to optimize IRS precodings. Zhu et al. \cite{g20} investigated the sum-rate maximization problem in an active IRS-assisted multi-user MISO system, which the iterative optimization approach was delvloped based on second-order cone programming and majorization-minimization methods. Based on the above analysis, active IRS has more advanced features than passive IRS. To futher improve the performance of communication system, we introduce active IRS and design active IRS and BS as an integrated system for joint control. In this case, it is natural to assume that active IRS will share a power supply with BS, and the optimal AR can be obtained by rationally distributing the power of BS and active IRS. Also this assumption facilitates performance comparisons with other systems.

The power allocation (PA) strategy can further improve rate performance under the condition that total power consumption of communication system is limited, which has been researched in \cite{g21,g22,g23,g24}. Specifically, a secrecy rate (SR) maximization problem was established and a PA strategy was proposed in a secure directional modulation system after specifying secrecy symbol and artificial noise beamformings \cite{g25}. The maximum SR of a secure spatial modulation system was investigated in \cite{g26} and two PA strategies were provided based on gradient descent method. The authors in \cite{g27} studied the total power consumption minimization problem in a cooperative downlink multi-user system and proposed a scheme for joint optimization of spectrum and PA factors. Two iterative algorithms, including inter-node and intra-node PA phases, were proposed for a full-duplex decode-and-forward MIMO relay system to improve the rate performance of users end \cite{g28}. 

The aforementioned research primarily centers on power distribution in wireless networks without IRS. Upon incorporating active IRS into the communication system, we investigated the potential rate improvement achievable by dynamically allocating power between the BS and active IRS, in contrast to traditional fixed PA. In this paper, our focus shifts to the development of two high-performance iterative PA strategies aimed at achieving corresponding PA gains. Our main contributions are outlined as follows:

\begin{itemize}
\item An active IRS-assisted PA wireless network system model is constructed and a PA strategy named Max-SNR-PA is proposed. We maximize SNR at user by jointly optimizing PA factor, BS transmit beamforming, and IRS phase shift matrix. Due to the fact that variables in objective function are coupled to each other, solving this problem directly is challenging. To deal with this difficulty, we adopt an alternate optimization approach. First, PA factor can be obtained by using polynomial regression method. Then, a successive convex approximation (SCA) technique in \cite{g29} is used to get BS transmit vector. Finally, a suboptimal iterative algorithm based on Dinkelbach's transform is applied for optimizing IRS phase shift beamforming.
\item To reduce the computational complexity of the Max-SNR-PA strategy,  a low-complexity Max-AR-CFFP algorithm is proposed. Here, we reformulate the system model with the goal of maximizing AR. Then, the objective function can be transformed into a equivalent form by using closed-form fractional programming (CFFP) method in \cite{g30,g31}. Next, we can use alternate iterative methods to obtain locally optimal solutions for BS and IRS beamformings. Simulation results show that: (a) the polynomial regression function can fit the original PA factor function well, (b) both of our proposed PA strategies can quickly achieve convergence, (c) compared with the case of fixed PA strategy in \cite{g32}, passive IRS, and without IRS, the two proposed PA strategies can effectively improve the rate performance.
\end{itemize}
 
The reminder of this paper is organized as follows. The system model with PA strategy is shown in Section 2. Two iterative PA algorithms are proposed in Section 3 and 4. Simulation results are presented in Section 5 and conclusions are drawn in Section 6.

\textit{Notations}: During this paper, matrices and vectors are denoted as uppercase letters and lowercase letters, respectively. $\mathbb{C}$ represents a set of complex numbers. $(\cdot)^H$, $(\cdot)^T$, $(\cdot)^*$, $||\cdot||$, $\rm diag(\cdot)$, $E\{\cdot\}$, and $\mathfrak{R}\{\cdot\}$ denote the conjugate transpose, transpose, conjugate, Euclidean norm, diagonal, expectation and real part operations, respectively. $\mathbf{I}_{N}$ represents the $N \times N$ identity matrix.

\rengSection{System Model} 
\begin{figure}
	\begin{center}
		\includegraphics[width=1.0\columnwidth,keepaspectratio]{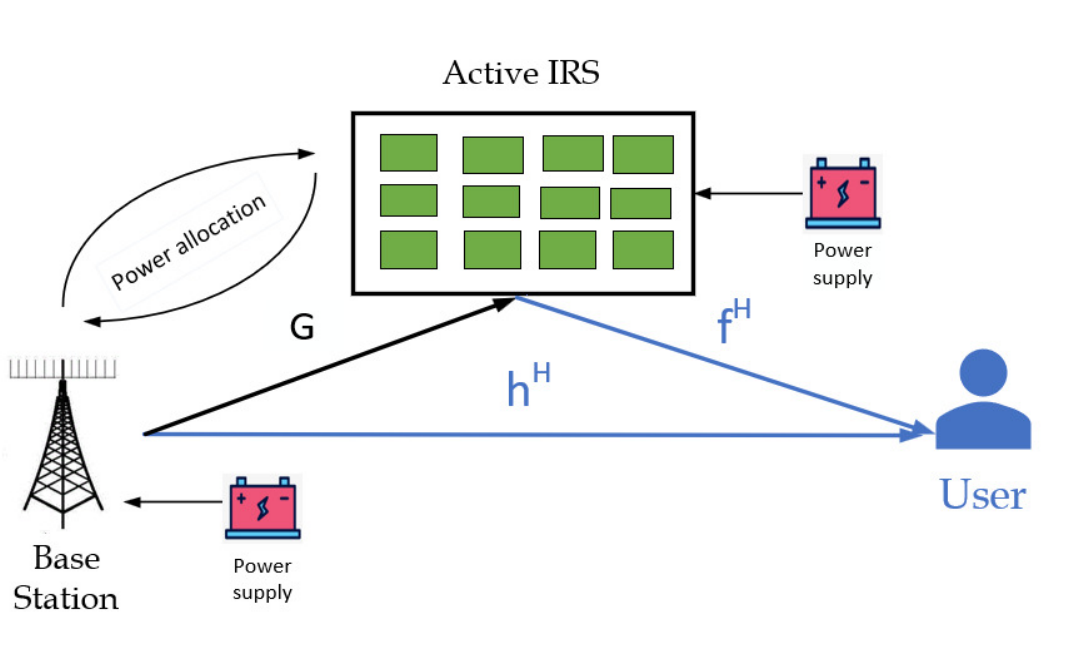}
		\fcaption{System model of an active IRS-assisted wireless network with PA.} 
	\end{center}
\end{figure}
\rengSubsection{System Model with PA factor}

Fig.1 shows an active IRS-assisted wireless network with PA. BS is equipped with $M$ antennas. The user is equipped with a single antenna, and active IRS is equipped with $N$ elements. $\mathbf{G}$ $\in$ $\mathbb{C}^{{N \times M}}$, $\mathbf{f}^H$ $\in$ $\mathbb{C}^{{1 \times N}}$, and $\mathbf{h}^H$ $\in$ $\mathbb{C}^{{1 \times M}}$ stand for the channels from BS to active IRS, active IRS to user, BS to user, respectively.
The transmitted signal at BS is given by
\begin{align}
	\mathbf{s}_B=\sqrt{\beta P_{max}}\mathbf{v}x,
\end{align}
where $\mathbf{v}$ $\in$ $\mathbb{C}^{{M \times 1}}$ is the transmit beamforming of BS that meets the condition $\mathbf{v}^H\mathbf{v}=1$, $P_{max}$ is the upper limit of the sum of power consumption by BS and active IRS, $\beta$ is the PA factor within interval [0,1], and $x$ denotes the transmit symbol satisfying $\mathbb{E}\left\{{|x|^2}\right\}=1$.

For active IRS, let $a_m$ and $\theta_m$ represent amplification coefficient of the $m$-th element and phase shift of m-th element respectively, where ${m = 1, \cdots, N}$. $\mathbf{\Theta}=\operatorname{diag}\left(a_1e^{j\theta_1},\cdots,a_Ne^{j\theta_N}\right)$ denotes the reflective beamforming matrix of active IRS. The signal reflected by active IRS can be written as:
\begin{align}
	\mathbf{s}_I=\mathbf{\Theta}\mathbf{G}\mathbf{s}_B+\mathbf{\Theta}\mathbf{n}_I
	=\sqrt{\beta P_{max}}\mathbf{\Theta}\mathbf{G}\mathbf{v}x+\mathbf{\Theta}\mathbf{n}_I,
\end{align}
where $\mathbf{n}_I$ $\in$ $\mathbb{C}^{{N \times 1}}$ denotes the additive white Gaussian noise (AWGN) introduced by active IRS power amplifiers, $\mathbf{n}_I \sim \mathcal{CN}\left( {0,{\mathbf{{\sigma}}_I^2\mathbf{I}_{N}}}\right)$. 

The received signal at user is given by
\begin{align}\label{3}
	y=\sqrt{\beta P_{max}}(\mathbf{f}^H\mathbf{\Theta}\mathbf{G}+\mathbf{h}^H)\mathbf{v}x+\mathbf{f}^H\mathbf{\Theta}\mathbf{n}_I+z,
\end{align}
where $z$ is the AWGN with distribution $z\sim\mathcal{C}\mathcal{N}(0,~\sigma^2_n)$. 

The power consumed at active IRS can be expressed as
	\begin{align}
		P_{IRS}&= E\{\mathbf{s}_I^H\mathbf{s}_I\}=\beta P_{max}||\mathbf{\Theta}\mathbf{G}\mathbf{v}||_2^2+\sigma^2_I||\mathbf{\Theta}||_F^2.
	\end{align}

From (3), the SNR at user can be given by
\begin{align}
	\text{SNR}
	&=\frac{\beta P_{max}||(\mathbf{f}^H\mathbf{\Theta}\mathbf{G}+\mathbf{h}^H)\mathbf{v}||_2^2}{\sigma^2_I||\mathbf{f}^H\mathbf{\Theta}||_2^2+\sigma^2_n}.
\end{align}

The achievable rate (AR) is
\begin{align}\label{7}
	\text{AR} =\rm log_2(1+\text{SNR}).
\end{align}
\rengSubsection{System Model without PA factor}

In this subsection, let us consider the system model hides PA factor $\beta$. 

The transmitted signal at BS is
\begin{align}
	\mathbf{s}_{B1}=\mathbf{v}_1x,
\end{align}
where $\mathbf{v}_1$ $\in$ $\mathbb{C}^{{M \times 1}}$ is the transmit beamforming of BS.

The signal reflected by active IRS is:
\begin{align}
	\mathbf{s}_{I1}=\mathbf{\Theta}\mathbf{G}\mathbf{s}_{B1}+\mathbf{\Theta}\mathbf{n}_I
	=\mathbf{\Theta}\mathbf{G}\mathbf{v}_1x+\mathbf{\Theta}\mathbf{n}_I.
\end{align}

The received signal at user can be given by
	\begin{align}
		y_1=(\mathbf{f}^H\mathbf{\Theta}\mathbf{G}+\mathbf{h}^H)\mathbf{v}_1x+\mathbf{f}^H\mathbf{\Theta}\mathbf{n}_I+z.
	\end{align}

The power consumed at BS and active IRS can be expressed as
\begin{equation}
	\begin{aligned}
		P_{BS1}&=\mathbf{v}_1^H\mathbf{v}_1,\\
		P_{IRS1}&=||\mathbf{\Theta}\mathbf{G}\mathbf{v}_1||_2^2+\sigma^2_I||\mathbf{\Theta}||_F^2.
	\end{aligned}
\end{equation}

The SNR is
\begin{align}
	\text{SNR}_1
	&=\frac{||(\mathbf{f}^H\mathbf{\Theta}\mathbf{G}+\mathbf{h}^H)\mathbf{v}_1||_2^2}{\sigma^2_I||\mathbf{f}^H\mathbf{\Theta}||_2^2+\sigma^2_n}.
\end{align}

The AR can be written as
\begin{align}\label{10}
	\text{AR}_1=\rm log_2(1+\text{SNR}_1).
\end{align}

\rengSection{Proposed Max-SNR-PA Strategy}

In this section, considering the system model with PA factor $\beta$, we maximize $\text{SNR}$ by jointly optimizing PA factor $\beta$, active IRS phase shift matrix $\mathbf{\Theta}$, and BS beamforming vector $\mathbf{v}$. The overall optimization problem is formulated as:
\begin{equation}
	\begin{aligned}
		\mathrm{(P0):}&\max_{\beta,\mathbf{\Theta},\mathbf{v}}~~~~~\text{SNR}
		=\frac{\beta P_{max}||(\mathbf{f}^H\mathbf{\Theta}\mathbf{G}+\mathbf{h}^H)\mathbf{v}||_2^2}{\sigma^2_I||\mathbf{f}^H\mathbf{\Theta}||_2^2+\sigma^2_n}\\
		&~~\text{s.t.}~~~~~~~0\le \beta \le 1,~ \mathbf{v}^H\mathbf{v}=1,\\
		&~~~~~~~~~~\beta P_{max}||\mathbf{\Theta}\mathbf{G}\mathbf{v}||_2^2+\sigma^2_I||\mathbf{\Theta}||_F^2 \le (1-\beta)P_{max}.
	\end{aligned}
\end{equation}

Due to the coupled $\beta$, $\mathbf{\Theta}$, and $\mathbf{v}$, this optimization problem is difficult to solve. In general, there is no efficient method to solve problem (P0) directly. Therefore, in the following, we apply the alternating optimization algorithm and optimize $\beta$, $\mathbf{\Theta}$, and $\mathbf{v}$ alternately.

\rengSubsection{Optimize $\beta$ by fixing $\mathbf{\Theta}$ and $\mathbf{v}$}

Letting $\boldsymbol{\theta}=(a_1e^{j\theta_1},\cdots,a_Ne^{j\theta_N})^H$, the received signal at user can be rewritten as
\begin{equation}
	\begin{split}
		y&=\sqrt{\beta P_{max}}(\rho\widetilde{\boldsymbol{\theta}}^H\operatorname{diag}(\mathbf{f}^H)\mathbf{G}+\mathbf{h}^H)\mathbf{v}x\\
		&+\rho\widetilde{\boldsymbol{\theta}}^H\operatorname{diag}(\mathbf{f}^H)\mathbf{n}_I+z,
	\end{split}
\end{equation}
where 
\begin{align}
	\boldsymbol{\theta}=\rho\widetilde{\boldsymbol{\theta}}, \rho=\|\boldsymbol{\theta}\|_2, \|\widetilde{\boldsymbol{\theta}}\|_2=1.
\end{align}

The power consumed at active IRS is
	\begin{align}
		P_{IRS}&=\beta P_{max}\rho^2||\widetilde{\boldsymbol{\theta}}^H\operatorname{diag}(\mathbf{G}\mathbf{v})||_2^2+\sigma^2_I\rho^2.
	\end{align}

In order to optimize $\beta$, we consider making the sum of consumed power of BS and active IRS reach the upper limit $P_{max}$, which means that
\begin{align}
	P_{IRS}=(1-\beta)P_{max},
\end{align}
then we can obtain
\begin{align}\label{1}
	\rho
	=\sqrt{\frac{(1-\beta)P_{max}}{\beta P_{max}||\widetilde{\boldsymbol{\theta}}^H\operatorname{diag}(\mathbf{G}\mathbf{v})||_2^2+\sigma^2_I}}.
\end{align}

Thus, the SNR can be simplified as
\begin{align}\label{2}
	\text{SNR}
	&=\frac{\beta P_{max}||(\rho\widetilde{\boldsymbol{\theta}}^H\operatorname{diag}(\mathbf{f}^H)\mathbf{G}+\mathbf{h}^H)\mathbf{v}||_2^2}{\sigma^2_I\rho^2||\widetilde{\boldsymbol{\theta}}^H\operatorname{diag}(\mathbf{f}^H)||_2^2+\sigma^2_n}.
\end{align}

Substituting (\ref{1}) into (\ref{2}) to simplify the optimization problem as
\begin{equation}
	\begin{aligned}
		\mathrm{(P1):}&\max_{\beta}~~~~~		
		f(\beta)=\frac{a\beta^2+b\beta+2c\beta\sqrt{d\beta^2+e\beta+f}}{g\beta+h}\\
		&~~\text{s.t.}~~~~~~~~0 \le \beta \le 1.
	\end{aligned}
\end{equation}
where

\begin{equation}
	\begin{aligned}
		a=&P_{max}^2||\mathbf{h}^H\mathbf{v}||^2||\widetilde{\boldsymbol{\theta}}^H\operatorname{diag}(\mathbf{G}\mathbf{v})||^2-\\
		&P_{max}^2||\widetilde{\boldsymbol{\theta}}^H\operatorname{diag}(\mathbf{f}^H)\mathbf{G}\mathbf{v}||^2,\\
		b=&P_{max}^2||\widetilde{\boldsymbol{\theta}}^H\operatorname{diag}(\mathbf{f}^H)\mathbf{G}\mathbf{v}||^2+P_{max}||\mathbf{h}^H\mathbf{v}||^2\sigma_I^2,\\
		c=&P_{max}\mathfrak{R}\{\widetilde{\boldsymbol{\theta}}^H\operatorname{diag}(\mathbf{f}^H)\mathbf{G}\mathbf{v}\mathbf{v}^H\mathbf{h}\},\\
		d=&-P_{max}^2||\widetilde{\boldsymbol{\theta}}^H\operatorname{diag}(\mathbf{G}\mathbf{v})||^2,\\
		e=&P_{max}^2||\widetilde{\boldsymbol{\theta}}^H\operatorname{diag}(\mathbf{G}\mathbf{v})||^2-\sigma_I^2P_{max},\\ 
		f=&P_{max}\sigma_I^2,\\
		g=&\sigma_n^2P_{max}||\widetilde{\boldsymbol{\theta}}^H\operatorname{diag}(\mathbf{G}\mathbf{v})||^2-\sigma_I^2P_{max}||\widetilde{\boldsymbol{\theta}}^H\operatorname{diag}(\mathbf{f}^H)||^2,\\
		h=&\sigma_I^2P_{max}||\widetilde{\boldsymbol{\theta}}^H\operatorname{diag}(\mathbf{f}^H)||^2+\sigma_n^2\sigma_I^2.
	\end{aligned}
\end{equation}

Due to the difficulty of directly solving the sub optimization problem (P1), in the following, we will use polynomial regression method to fit the objective function $f(\beta)$. Firstly, a $Q$-order polynomial $g(\beta)$ is constructed to approximate $f(\beta)$,
\begin{equation}
	\begin{aligned}
		g(\beta)&=a_0+a_1\beta+\cdots+a_Q\beta^Q\\
		&=\left(1,\beta,\cdots,\beta^Q\right)\bullet\left(a_0,a_1,\cdots,a_Q\right)^T.
	\end{aligned}
\end{equation}

 To estimate the coefficients of the above polynomial, the training set $S_{PR}$ is generated as
\begin{align}
	S_{PR}=\left\{(\beta_1, f(\beta_1)), (\beta_2, f(\beta_2)), \cdots, (\beta_{J}, f(\beta_{J}))\right\}.
\end{align}

Constructing the polynomial fitting matrix-vector form as follows:
\begin{align}
	\underbrace{\begin{pmatrix}
			g(\beta_1)\\
			\vdots\\
			g(\beta_{J})
	\end{pmatrix}}_{\mathbf{g}}=
	\underbrace{\begin{pmatrix}
			1 & \beta_1 & \cdots & \beta_1^Q\\
			\vdots & \vdots & \vdots & \vdots\\
			1 & \beta_{J} & \cdots & \beta_{J}^Q
	\end{pmatrix}}_{\mathbf{A}}
	\underbrace{\begin{pmatrix}
			a_0\\
			\vdots\\
			a_Q
	\end{pmatrix}}_{\boldsymbol{b}},
\end{align}
$\mathbf{g}=\mathbf{A}\boldsymbol{b}$, where $J \geq 5(Q+1)$. Let us define the following target vector 
\begin{align}
	\mathbf{c}=\left(f(\beta_1), \cdots, f(\beta_J)\right)^T.
\end{align}

The corresponding square error summation is defined as
\begin{equation}
	\begin{aligned}
		\Delta(\boldsymbol{b})&=(\mathbf{g}-\mathbf{c})^T(\mathbf{g}-\mathbf{c})/J\\
		&=\frac{1}{J}\left\{\boldsymbol{b}^T\mathbf{A}^T\mathbf{A}\boldsymbol{b}-\mathbf{c}^T\mathbf{A}\boldsymbol{b}-\boldsymbol{b}^T\mathbf{A}^T\mathbf{c}+\mathbf{c}^T\mathbf{c}\right\}.
	\end{aligned}
\end{equation}

Taking the first derivative of $\Delta(\boldsymbol{b})$ with respect to $\boldsymbol{b}$ equal zero,
\begin{align}
	\frac{\partial\Delta(\boldsymbol{b})}{\partial \boldsymbol{b}}=\frac{1}{J}\left\{2\mathbf{A}^T\mathbf{A}\boldsymbol{b}-2\mathbf{A}^T\mathbf{c}\right\}=0,
\end{align}
which yields
\begin{align}
	\hat{\boldsymbol{b}}=\left(\mathbf{A}^T\mathbf{A}\right)^{-1}\mathbf{A}^T\mathbf{c}.
\end{align}

We have completed the esitmate of coefficients of polynomial $g(\beta)$ and obtained the fitting polynomial as follows:
\begin{align}
	\hat{g}(\beta)&=\hat{a}_0+\hat{a}_1\beta+\cdots+\hat{a}_Q\beta^Q.
\end{align}

Finding the stationary points of the above polynomial in interval [0,1] is eqivuelent to find the roots of the following polynomial:
\begin{align}\label{Root-Equ-M}
	\frac{\partial\hat{g}(\beta)}{\partial\beta}=Q\hat{a}_Q\beta^{Q-1}+\cdots+2\hat{a}_2\beta+\hat{a}_1=0.
\end{align}

For example, when $Q=2$, we have 
\begin{align}
	2\hat{a}_2\beta+\hat{a}_1=0 \Rightarrow \bar{\beta}=\frac{-\hat{a}_1}{2\hat{a}_2},
\end{align}
when $Q=3$, we have 
\begin{align}
	3\hat{a}_3\beta^2+2\hat{a}_2\beta+\hat{a}_1=0,
\end{align}
which yields
\begin{align}
	\bar{\beta}_1=\frac{-\hat{a}_2+\sqrt{\hat{a}_2^2-3\hat{a}_3\hat{a}_1}}{3\hat{a}_3}, \bar{\beta}_2=\frac{-\hat{a}_2-\sqrt{\hat{a}_2^2-3\hat{a}_3\hat{a}_1}}{3\hat{a}_3}.
\end{align}

In order to guarantee that there are closed-form roots for the equation (\ref{Root-Equ-M}), the value of $Q$ is taken to be an integer smaller than 6. 

Considering $\beta \in [0,1]$, we need to judge whether all candidate roots are within the interval [0,1], then we have
\begin{equation}
	\begin{aligned}
		&\widetilde{\beta_i}=\begin{cases}
			\bar{\beta_i}, \bar{\beta_i} \in [0,1],\\
			0, ~\bar{\beta_i} \notin [0,1].
		\end{cases}
	\end{aligned}
\end{equation}
where $i\in{\left\{1,2,\cdots,Q-1\right\}}$, the optimal solution of $\beta$ is

\begin{align}
	\beta^{o}=\mathop{\rm argmax} \limits_{\beta \in S_B} ~f(\beta),
\end{align}
where set $S_B$ is defined as
\begin{align}
	S_B=\left\{0, \widetilde{\beta}_1, \cdots, \widetilde{\beta}_{Q-1}, 1\right\}.
\end{align}

\rengSubsection{Optimize $\mathbf{v}$ by fixing $\beta$ and $\mathbf{\Theta}$}

In this subsection, beamforming vector of BS $\mathbf{v}$ is optimized by fixing PA factor $\beta$ and active IRS phase shift matrix $\mathbf{\Theta}$. The optimization problem with respect to $\mathbf{v}$ is
\begin{equation}
	\begin{aligned}
		\mathrm{(P2):}&\max_{\mathbf{v}}~~~~~\text{SNR}
		=\frac{\beta P_{max}||(\mathbf{f}^H\mathbf{\Theta}\mathbf{G}+\mathbf{h}^H)\mathbf{v}||_2^2}{\sigma^2_I||\mathbf{f}^H\mathbf{\Theta}||_2^2+\sigma^2_n}\\
		&~~\text{s.t.}~~~~~~~~ \mathbf{v}^H\mathbf{v}=1,\\
		&~~~~~~~~~~\beta P_{max}||\mathbf{\Theta}\mathbf{G}\mathbf{v}||_2^2+\sigma^2_I||\mathbf{\Theta}||_F^2 \le (1-\beta)P_{max}.
	\end{aligned}
\end{equation}
which can be re-arranged as

\begin{equation}
	\begin{aligned}
		\mathrm{(P2-1):}&\max_{\mathbf{v}}~~~~~~~
		\mathbf{v}^H\mathbf{B}\mathbf{v}\\
		&~~\text{s.t.}~~~~~~~~ \mathbf{v}^H\mathbf{v}=1,\\
		&~~~~~~~~~~\mathbf{v}^H\mathbf{C}\mathbf{v} \le P'_{max}.
	\end{aligned}
\end{equation}
where

\begin{equation}
	\begin{aligned}
		&\mathbf{B}=(\mathbf{f}^H\mathbf{\Theta}\mathbf{G}+\mathbf{h}^H)^H(\mathbf{f}^H\mathbf{\Theta}\mathbf{G}+\mathbf{h}^H),\\
		&\mathbf{C}=\mathbf{G}^H\mathbf{\Theta}^H\mathbf{\Theta}\mathbf{G},\\
		&P'_{max}=\frac{(1-\beta)P_{max}-\sigma^2_I||\mathbf{\Theta}||_F^2}{\beta P_{max}}.
	\end{aligned}
\end{equation}

Due to the insensitivity of the objective function value to the scaling of $\mathbf{v}$, we relax the modulo constraint to $\mathbf{v}^H\mathbf{v} \le 1$, the optimization problem can be rewritten as
\begin{equation}
	\begin{aligned}
		\mathrm{(P2-2):}&\max_{\mathbf{v}}~~~~~~~
		\mathbf{v}^H\mathbf{B}\mathbf{v}\\
		&~~\text{s.t.}~~~~~~~~ \mathbf{v}^H\mathbf{v} \le 1,\\
		&~~~~~~~~~~\frac{\mathbf{v}^H\mathbf{C}\mathbf{v}}{\mathbf{v}^H\mathbf{v}} \le P'_{max}.
	\end{aligned}
\end{equation}

However, the optimization problem (P2-2) is still non-convex. Therefore, we use SCA method to solve this problem. By referring to the first-order Taylor series expansion at fixed point $\tilde{\mathbf{v}}$, we have
\begin{equation}
	\begin{aligned}
		&\mathbf{v}^H\mathbf{B}\mathbf{v} \ge 2\mathfrak{R}\{\tilde{\mathbf{v}}^H\mathbf{B}\mathbf{v}\}-\tilde{\mathbf{v}}^H\mathbf{B}\tilde{\mathbf{v}},\\
		&\mathbf{v}^H\mathbf{v} \ge 2\mathfrak{R}\{\tilde{\mathbf{v}}^H\mathbf{v}\}-\tilde{\mathbf{v}}^H\tilde{\mathbf{v}}.
	\end{aligned}
\end{equation}
then, the optimization problem is transformed into

\begin{equation}\label{8}
	\begin{aligned}
		\mathrm{(P2-3):}&\max_{\mathbf{v}}~~~~~~~
		\mathfrak{R}\{\tilde{\mathbf{v}}^H\mathbf{B}\mathbf{v}\}\\
		&~~\text{s.t.}~~~~~~~~ \mathbf{v}^H\mathbf{v} \le 1,\\
		&~~~~~~~~~~\mathbf{v}^H\mathbf{C}\mathbf{v} \le P'_{max}\left(2\mathfrak{R}\{\tilde{\mathbf{v}}^H\mathbf{v}\}-\tilde{\mathbf{v}}^H\tilde{\mathbf{v}}\right).
	\end{aligned}
\end{equation}

This is a convex optimization problem, and it can be solved by CVX. After obtain the solution $\bar{\mathbf{v}}$, the beamforming vector is designed as
\begin{align}
	\mathbf{v}=\frac{\bar{\mathbf{v}}}{|\bar{\mathbf{v}}|}.
\end{align}

\rengSubsection{Optimize $\mathbf{\Theta}$ by fixing $\beta$ and $\mathbf{v}$}

In this subsection, we optimize $\mathbf{\Theta}$ by fixing $\beta$ and $\mathbf{v}$. Considering reflective beamforming vector $\boldsymbol{\theta}$, and on the basis of (\ref{3}), the received signal at user can be rewritten as
\begin{equation}
	\begin{aligned}
		y&=\sqrt{\beta P_{max}}(\boldsymbol{\theta}^H\operatorname{diag}(\mathbf{f}^H)\mathbf{G}+\mathbf{h}^H)\mathbf{v}x\\
		&+\boldsymbol{\theta}^H\operatorname{diag}(\mathbf{f}^H)\mathbf{n}_I+z,
	\end{aligned}
\end{equation}
SNR is given by

\begin{align}
	\text{SNR}
	&=\frac{\beta P_{max}||(\boldsymbol{\theta}^H\operatorname{diag}(\mathbf{f}^H)\mathbf{G}+\mathbf{h}^H)\mathbf{v}||_2^2}{\sigma^2_I||\boldsymbol{\theta}^H\operatorname{diag}(\mathbf{f}^H)||_2^2+\sigma^2_n}.
\end{align}

The power consumed at active IRS can be reformulated as
\begin{equation}
	\begin{aligned}
		&P_{IRS}=\beta P_{max}||\boldsymbol{\theta}^H\operatorname{diag}(\mathbf{G}\mathbf{v})||_2^2+\sigma^2_I\boldsymbol{\theta}^H\boldsymbol{\theta}\\
		&=\beta P_{max}\boldsymbol{\theta}^H\operatorname{diag}(\mathbf{G}\mathbf{v})\operatorname{diag}(\mathbf{G}\mathbf{v})^H\boldsymbol{\theta}+\sigma^2_I\boldsymbol{\theta}^H\boldsymbol{\theta}\\
		&=\boldsymbol{\theta}^H\left[\beta P_{max}\operatorname{diag}(\mathbf{G}\mathbf{v})\operatorname{diag}(\mathbf{G}\mathbf{v})^H+\sigma^2_I\mathbf{I}_{N}\right]\boldsymbol{\theta}.
	\end{aligned}
\end{equation}

Simplifying the numerator and denominator terms in SNR can yield
\begin{equation}
	\begin{aligned}
		&||(\boldsymbol{\theta}^H\operatorname{diag}(\mathbf{f}^H)\mathbf{G}+\mathbf{h}^H)\mathbf{v}||_2^2\\
		&=(\boldsymbol{\theta}^H\operatorname{diag}(\mathbf{f}^H)\mathbf{G}+\mathbf{h}^H)\mathbf{v}\mathbf{v}^H(\boldsymbol{\theta}^H\operatorname{diag}(\mathbf{f}^H)\mathbf{G}+\mathbf{h}^H)^H\\
		&=(\boldsymbol{\theta}^H\operatorname{diag}(\mathbf{f}^H)\mathbf{G}+\mathbf{h}^H)\mathbf{v}\mathbf{v}^H(\mathbf{G}^H\operatorname{diag}(\mathbf{f}^H)^H\boldsymbol{\theta}+\mathbf{h})\\
		&=\boldsymbol{\theta}^H\operatorname{diag}(\mathbf{f}^H)\mathbf{G}\mathbf{v}\mathbf{v}^H\mathbf{G}^H\operatorname{diag}(\mathbf{f}^H)^H\boldsymbol{\theta}+\mathbf{h}^H\mathbf{v}\mathbf{v}^H\mathbf{h}\\
		&+2\mathfrak{R}\{\mathbf{h}^H\mathbf{v}\mathbf{v}^H\mathbf{G}^H\operatorname{diag}(\mathbf{f}^H)^H\boldsymbol{\theta}\},\\
		&\sigma^2_I||\boldsymbol{\theta}^H\operatorname{diag}(\mathbf{f}^H)||_2^2+\sigma^2_n=\sigma^2_I\boldsymbol{\theta}^H\operatorname{diag}(\mathbf{f}^H)\operatorname{diag}(\mathbf{f}^H)^H\boldsymbol{\theta}+\sigma^2_n.
	\end{aligned}
\end{equation}

Thus, the optimization problem respect to $\boldsymbol{\theta}$ can be recast as
\begin{equation}
	\begin{aligned}
		\mathrm{(P3):}&\max_{\boldsymbol{\theta}}~~~~~~~
		\frac{\boldsymbol{\theta}^H\mathbf{D}\boldsymbol{\theta}+2\mathfrak{R}\{\mathbf{t}^H\boldsymbol{\theta}\}}{\boldsymbol{\theta}^H\mathbf{E}\boldsymbol{\theta}+\sigma^2_n}\\
		&~~\text{s.t.}~~~~~~~~\boldsymbol{\theta}^H\mathbf{F}\boldsymbol{\theta} \le (1-\beta)P_{max}.
	\end{aligned}
\end{equation}
where for briefly, we define

\begin{equation}
	\begin{aligned}
		\mathbf{D}=&\operatorname{diag}(\mathbf{f}^H)\mathbf{G}\mathbf{v}\mathbf{v}^H\mathbf{G}^H\operatorname{diag}(\mathbf{f}^H)^H,\\
		\mathbf{t}^H=&\mathbf{h}^H\mathbf{v}\mathbf{v}^H\mathbf{G}^H\operatorname{diag}(\mathbf{f}^H)^H,\\
		\mathbf{E}=&\sigma^2_I\operatorname{diag}(\mathbf{f}^H)\operatorname{diag}(\mathbf{f}^H)^H,\\
		\mathbf{F}=&\beta P_{max}\operatorname{diag}(\mathbf{G}\mathbf{v})\operatorname{diag}(\mathbf{G}\mathbf{v})^H+\sigma^2_I\mathbf{I}_{N}.
	\end{aligned}
\end{equation}

Since this is a fractional programming problem, we can use the Dinkelbach's transform, then the optimization problem turns into
\begin{equation}
	\begin{aligned}
		\mathrm{(P3-1):}&\max_{\boldsymbol{\theta}}~~~~~~~
		\boldsymbol{\theta}^H\mathbf{D}\boldsymbol{\theta}+2\mathfrak{R}\{\mathbf{t}^H\boldsymbol{\theta}\}-\eta(\boldsymbol{\theta}^H\mathbf{E}\boldsymbol{\theta}+\sigma^2_n)\\
		&~~\text{s.t.}~~~~~~~~\boldsymbol{\theta}^H\mathbf{F}\boldsymbol{\theta} \le (1-\beta)P_{max}.
	\end{aligned}
\end{equation}
where $\eta$ is an auxiliary variable, during each iteration process

\begin{align}
	\eta^{(i+1)}=\frac{\boldsymbol{\theta}^{(i)H}\mathbf{D}\boldsymbol{\theta}^{(i)}+2\mathfrak{R}\{\mathbf{t}^H\boldsymbol{\theta}^{(i)}\}}{\boldsymbol{\theta}^{(i)H}\mathbf{E}\boldsymbol{\theta}^{(i)}+\sigma^2_n}.
\end{align}

Similarly, by using the first-order Taylor series expansion at fixed point  $\boldsymbol{\theta}_0$, we have
\begin{align}
	\boldsymbol{\theta}^H\mathbf{D}\boldsymbol{\theta} \ge 2\mathfrak{R}\{\boldsymbol{\theta}_0^H\mathbf{D}\boldsymbol{\theta}\}-\boldsymbol{\theta}_0^H\mathbf{D}\boldsymbol{\theta}_0,
\end{align}
then, problem (P3-1) can be transformed into
\begin{equation}\label{4}
	\begin{aligned}
		\mathrm{(P3-2):}&\max_{\boldsymbol{\theta}}~~~~~~~
		2\mathfrak{R}\{\boldsymbol{\theta}_0^H\mathbf{D}\boldsymbol{\theta}\}+2\mathfrak{R}\{\mathbf{t}^H\boldsymbol{\theta}\}-\eta(\boldsymbol{\theta}^H\mathbf{E}\boldsymbol{\theta}+\sigma^2_n)\\
		&~~\text{s.t.}~~~~~~~~\boldsymbol{\theta}^H\mathbf{F}\boldsymbol{\theta} \le (1-\beta)P_{max}.
	\end{aligned}
\end{equation}

The optimization problem (\ref{4}) is convex, and we can address it by CVX directly. The process of algorithm to optimize $\mathbf{\Theta}$ is as follows:

\noindent\begin{tabular*}{\columnwidth}{l}
	\hline
	{\bf Algorithm 1.} The algorithm to optimize $\mathbf{\Theta}$\\
	\hline
\end{tabular*}
\begin{algorithmic}
	\STATE  1: Input $\mathbf{D}$, $\mathbf{t}^H$, $\mathbf{E}$, $\mathbf{F}$, initial value $\boldsymbol{\theta}^{(0)}$, $\eta^{(0)}$, $i=0$, and convergence accuracy $\xi$.
	\REPEAT 
	\STATE  2: $i=i+1$.
	\STATE 3: Update $\eta^{(i)}$, $\eta^{(i)}=\frac{\boldsymbol{\theta}^{((i-1))H}\mathbf{D}\boldsymbol{\theta}^{(i-1)}+2\mathfrak{R}\{\mathbf{t}^H\boldsymbol{\theta}^{(i-1)}\}}{\boldsymbol{\theta}^{((i-1))H}\mathbf{E}\boldsymbol{\theta}^{(i-1)}+\sigma^2_n}$.
	\STATE 4: Let $\boldsymbol{\theta}_0=\boldsymbol{\theta}^{(i-1)}$, solved problem (\ref{4}) to obtain $\boldsymbol{\theta}^{(i)}$.
	\UNTIL  $\boldsymbol{\theta}^{(i)H}\mathbf{D}\boldsymbol{\theta}^{(i)}+2\mathfrak{R}\{\mathbf{t}^H\boldsymbol{\theta}^{(i)}\}-\eta^{(i)}(\boldsymbol{\theta}^{(i)H}\mathbf{E}\boldsymbol{\theta}^{(i)}+\sigma^2_n) \le \xi$.
	\STATE 5: Output $\boldsymbol{\theta}$, $\mathbf{\Theta}=\operatorname{diag}(\boldsymbol{\theta}^H)$.
\end{algorithmic}
\begin{tabular*}{\columnwidth}{l}
	\hline
\end{tabular*}

\rengSubsection{Overall strategy and complexity analysis}

In this subsection, we have summarized the algorithm implementation of alternatingly optimizing variables $\beta$, $\mathbf{\Theta}$, and $\mathbf{v}$ as follows:

\noindent\begin{tabular*}{\columnwidth}{l}
	\hline
	{\bf Algorithm 2.} Proposed Max-SNR-PA algorithm\\
	\hline
\end{tabular*}
\begin{algorithmic}
	\STATE  1: Initialize feasible solutions $\beta^{(0)}$, $\mathbf{v}^{(0)}$, and $\mathbf{\Theta}^{(0)}$, calculate the achievable rate $\text{AR}^{(0)}$ based on (\ref{7}).
	\STATE 2: Set the iteration number $k=0$, convergence accuracy $\varepsilon$.
	\REPEAT
	\STATE  3: Given $\mathbf{\Theta}^{(k)}$ and $\mathbf{v}^{(k)}$ to obtain $\beta^{(k+1)}$ based on (\ref{Root-Equ-M}).
	\STATE 4: Given $\mathbf{\Theta}^{(k)}$ and $\beta^{(k+1)}$ to obtain $\mathbf{v}^{(k+1)}$ based on (\ref{8}).
	\STATE 5: Given $\mathbf{v}^{(k+1)}$ and $\beta^{(k+1)}$ to obtain $\mathbf{\Theta}^{(k+1)}$ based on (\ref{4}).
	\STATE 6: $k=k+1$.
	\UNTIL $|\text{AR}^{(k)}-\text{AR}^{(k-1)}| \le \varepsilon$.
	\STATE 7: $\mathbf{v}^{(k)}$, $\beta^{(k)}$, and $\mathbf{\Theta}^{(k)}$ are the optimal value, and $\text{AR}^{(k)}$ is the optimal achievable rate.
\end{algorithmic}
\begin{tabular*}{\columnwidth}{l}
	\hline
\end{tabular*}

Due to the fact that the obtained solutions in Algorithm 2 are locally optimal, and the objective value sequence $\left\{\text{AR}(\beta^{(k)}, \mathbf{v}^{(k)}, \mathbf{\Theta}^{(k)})\right\}$ obtained in each iteration of the alternate optimization method is non-decreasing. Specifically, it follows
\begin{equation}
	\begin{aligned}
		&\text{AR}\left(\beta^{(k)}, \mathbf{v}^{(k)}, \mathbf{\Theta}^{(k)}\right)\\
		&\stackrel{(a)}{\le} \text{AR}\left(\beta^{(k+1)}, \mathbf{v}^{(k)}, \mathbf{\Theta}^{(k)}\right)\\
		&\stackrel{(b)}{\le} \text{AR}\left(\beta^{(k+1)}, \mathbf{v}^{(k+1)}, \mathbf{\Theta}^{(k)}\right)\\
		&\stackrel{(c)}{\le} \text{AR}\left(\beta^{(k+1)}, \mathbf{v}^{(k+1)}, \mathbf{\Theta}^{(k+1)}\right),
	\end{aligned}
\end{equation}

where $(a)$, $(b)$, and $(c)$ are due to the update in $(\ref{Root-Equ-M})$, $(\ref{8})$, and $(\ref{4})$, respectively. Moreover, $\text{AR}\left(\beta^{(k)}, \mathbf{v}^{(k)}, \mathbf{\Theta}^{(k)}\right)$ has a finite upper bound since the limited power constraint. Therefore, the convergence of proposed Max-SNR-PA algorithm can be guaranteed.

The computational complexity of \textbf{Algorithm 2} is mainly determined by the updates of the three variables $\beta$, $\mathbf{v}$, and $\mathbf{\Theta}$ via $(\ref{Root-Equ-M})$, $(\ref{8})$, and $(\ref{4})$, respectively. Specifically, the computational complexity of updating $\beta$ is $\mathcal{O}\left\{(Q+1)^4\right\}$ float-point operations (FLOPs). The complexity of updating $\mathbf{v}$ is $\mathcal{O}\left\{6M^3\rm log_2(1/\varepsilon)\right\}$ FLOPs. The complexity of updating $\mathbf{\Theta}$ is $\mathcal{O}\left\{{\rm log_2(1/\varepsilon)} L_{\mathbf{\Theta}}{\rm log_2(1/\xi)}\sqrt{N}(2N^4+N^3)\right\}$ FLOPs. Thus, the overall computational complexity of \textbf{Algorithm 2} is given by $\mathcal{O}\left\{L_p[Q^4+6M^3{\rm log_2(1/\varepsilon)}+2{\rm log_2(1/\varepsilon)}L_{\mathbf{\Theta}}{\rm log_2(1/\xi)}N^{4.5}]\right\}$, wherein $Q$ is the order of fitting polynomial, $\xi$ is the given accuracy tolerance of \textbf{Algorithm 1}, $\varepsilon$ is the given accuracy tolerance of \textbf{Algorithm 2}, $L_{\mathbf{\Theta}}$ denotes the number of iterations required by \textbf{Algorithm 1} for convergence, $L_p$ denotes the number of iterations required by \textbf{Algorithm 2} for convergence.

\rengSection{Proposed Max-AR-CFFP Strategy}

	In the previous section, we have proposed a PA strategy named Max-AR-PA to achieve the power allocation between BS and active IRS. However, the computational complexity of this algorithm is too high. To address this issue, a low-complexity alternating iteration method will be presented as follows. This method differs by optimizing the beamforming vector of the BS and the active IRS phase shift matrix, with the optimization goal being the AR rather than the SNR. Thus, let us consider the system model hiding PA factor $\beta$ mentioned in subsection 2.2.

Our optimization goal is to maximize $\text{AR}_1$ by jointly optimizing $\mathbf{v}_1$ and $\mathbf{\Theta}$ under limited total power $P_{max}$. The overall optimization problem is formulated as:
\begin{equation}
	\begin{aligned}
		\mathrm{(P4):}&\max_{\mathbf{\Theta},\mathbf{v}_1}~~~~~~~~~~~~~\text{AR} _1\\
		&~~\text{s.t.}~~~~~~~P_{BS1}+P_{IRS1} \le P_{max}.
	\end{aligned}
\end{equation}

The joint design of $\mathbf{\Theta}$ and $\mathbf{v}_1$ is challenging due to the non-convexity and highly coupled variables in problem (P4). Therefore, in order to effectively solve this problem, we develop a joint beamforming and precoding scheme based on alternate optimization and closed-form fractional programming method. Introducing auxiliary variables $\gamma$ and $\mu$, the original optimization problem (P4) is equivalent to
\begin{equation}\label{5}
	\begin{aligned}
		\mathrm{(P4-1):}&\max_{\mathbf{\Theta},\mathbf{v}_1, \gamma, \mu}~\text{AR}_1'=\rm ln(1+\gamma)-|\mu|^2(\sigma^2_I||\mathbf{f}^H\mathbf{\Theta}||_2^2+\sigma^2_n)\\
		&-\gamma+2\sqrt{(1+\gamma)}\mathfrak{R}\{\mu^*(\mathbf{f}^H\mathbf{\Theta}\mathbf{G}+\mathbf{h}^H)\mathbf{v}_1\}\\
		&~~\text{s.t.}~~~~~~~\mathbf{v}_1^H\mathbf{v}_1+||\mathbf{\Theta}\mathbf{G}\mathbf{v}_1||_2^2+\sigma^2_I||\mathbf{\Theta}||_F^2 \le P_{max}.
	\end{aligned}
\end{equation}

Then, the locally optimal solution of (\ref{5}) can be obtained by optimizing these variables alternately.

\rengSubsection{Optimize $\mu$, given $\mathbf{v}_1$, $\mathbf{\Theta}$, and $\gamma$}

After giving $\mathbf{v}_1$, $\mathbf{\Theta}$, and $\gamma$, the optimal $\mu$ can be obtained by solving $\frac{\partial \rm AR_1'}{\partial \mu}=0$ as
\begin{align}\label{11}
	\mu^{opt}=\frac{\sqrt{(1+\gamma)||(\mathbf{f}^H\mathbf{\Theta}\mathbf{G}+\mathbf{h}^H)\mathbf{v}_1||_2^2}}{\sigma^2_I||\mathbf{f}^H\mathbf{\Theta}||_2^2+\sigma^2_n}.
\end{align}

\rengSubsection{Optimize $\gamma$, given $\mathbf{v}_1$, $\mathbf{\Theta}$, and $\mu$}

After giving $\mathbf{v}_1$, $\mathbf{\Theta}$, and $\mu$, the optimal $\gamma$ can be obtained by solving $\frac{\partial \rm AR_1'}{\partial \gamma}=0$ as
\begin{align}\label{12}
	\gamma^{opt}=\frac{\varpi^2+\varpi\sqrt{\varpi^2+4}}{2},
\end{align}
where $\varpi=\mathfrak{R}\{\mu^*(\mathbf{f}^H\mathbf{\Theta}\mathbf{G}+\mathbf{h}^H)\mathbf{v}_1\}$.

\rengSubsection{Optimize $\mathbf{v}_1$, given $\mathbf{\Theta}$, $\gamma$, and $\mu$}

For briefly, we define
\begin{equation}
	\begin{aligned}
		\mathbf{H}=&\mathbf{I}_{M}+\mathbf{G}^H\mathbf{\Theta}^H\mathbf{\Theta}\mathbf{G},\\
		\mathbf{k}^H=&2\sqrt{(1+\gamma)}\mu^*(\mathbf{f}^H\mathbf{\Theta}\mathbf{G}+\mathbf{h}^H),\\
		P_r=&P_{max}-\sigma^2_I||\mathbf{\Theta}||_F^2.
	\end{aligned}
\end{equation}
then, problem (P4-1) can be reformulated as follows

\begin{equation}\label{13}
	\begin{aligned}
		\mathrm{(P4-2):}&\max_{\mathbf{v}_1}~~~~~\mathfrak{R}\{\mathbf{k}^H\mathbf{v}_1\}\\
		&~~\text{s.t.}~~~~~~~\mathbf{v}_1^H\mathbf{H}\mathbf{v}_1 \le P_r.
	\end{aligned}
\end{equation}
this is a convex problem and it can be solved by CVX.

\rengSubsection{Optimize $\mathbf{\Theta}$, given $\mathbf{v}_1$, $\gamma$, and $\mu$}

Before solving this optimization problem, we have
\begin{equation}\label{9}
	\begin{aligned}
		\mathbf{f}^H\mathbf{\Theta}\mathbf{G}+\mathbf{h}^H&=\boldsymbol{\theta}^H\operatorname{diag}(\mathbf{f}^H)\mathbf{G}+\mathbf{h}^H,\\
		||\mathbf{\Theta}\mathbf{G}\mathbf{v}_1||_2^2&=||\boldsymbol{\theta}^H\operatorname{diag}(\mathbf{G}\mathbf{v}_1)||_2^2,\\
		||\mathbf{\Theta}||_F^2&=\boldsymbol{\theta}^H\boldsymbol{\theta}.
	\end{aligned}
\end{equation}

Utilizing (\ref{9}), while giving $\mathbf{v}_1$, $\gamma$ and $\mu$, problem (P4-1) can be reformulated as
\begin{equation}\label{14}
	\begin{aligned}
		\mathrm{(P4-3):}&\max_{\boldsymbol{\theta}}~~~~~\mathfrak{R}\{\boldsymbol{\theta}^H\mathbf{d}\}-\boldsymbol{\theta}^H\mathbf{J}\boldsymbol{\theta}\\
		&~~\text{s.t.}~~~~~~~\boldsymbol{\theta}^H\mathbf{L}\boldsymbol{\theta} \le P_b.
	\end{aligned}
\end{equation}
where
\begin{equation}
	\begin{aligned}
		\mathbf{d}&=2\sqrt{(1+\gamma)}\operatorname{diag}(\mu^*\mathbf{f}^H)\mathbf{G}\mathbf{v}_1,\\
		\mathbf{J}&=|\mu|^2\sigma^2_I\operatorname{diag}(\mathbf{f}^H)\operatorname{diag}(\mathbf{f}^H)^H,\\
		\mathbf{L}&=\operatorname{diag}(\mathbf{G}\mathbf{v}_1)\operatorname{diag}(\mathbf{G}\mathbf{v}_1)^H+\sigma_I^2\mathbf{I}_{N},\\
		P_b&=P_{max}-\mathbf{v}_1^H\mathbf{v}_1.
	\end{aligned}
\end{equation}

This is a convex problem, it could be solved by CVX. Then, we can obtain
 $\mathbf{\Theta}=\operatorname{diag}(\boldsymbol{\theta}^H)$.
 
 \rengSubsection{Overall strategy and complexity analysis}
 
 In this sunsection, we have summarized the algorithm implementation process for alternating optimization variables $\mathbf{v}_1$, $\mathbf{\Theta}$, $\mu$, and $\gamma$ as follows:
 
 \noindent\begin{tabular*}{\columnwidth}{l}
 	\hline
 	{\bf Algorithm 3.} Proposed Max-AR-CFFP algorithm\\
 	\hline
 \end{tabular*}
 \begin{algorithmic}
 	\STATE  1: Initialize $\mathbf{v}_1^{(0)}$, $\mathbf{\Theta}^{(0)}$, $\mu^{(0)}$, and $\gamma^{(0)}$, calculate the achievable rate $\text{AR}_1^{(0)}$ based on (\ref{10}).
 	\STATE 2: Set $t=0$, convergence accuracy $\zeta$.
 	\REPEAT
 	\STATE 2: Update $\mu^{(t+1)}$ by (\ref{11}).
 	\STATE 3: Update $\gamma^{(t+1)}$ by (\ref{12}).
 	\STATE 4: Update $\mathbf{v}_1^{(t+1)}$ by (\ref{13}).
 	\STATE 5: Update $\mathbf{\Theta}^{(t+1)}$ by (\ref{14}).
 	\STATE 6: $t=t+1$.
 	\UNTIL $|\text{AR}_1^{(t)}-\text{AR}_1^{(t-1)}| \le \zeta$.
 	\STATE 7: $\mathbf{v}_1$ and $\mathbf{\Theta}$ are the optimal value, and $\text{AR}_1$ is the optimal achievable rate.
 \end{algorithmic}
 \begin{tabular*}{\columnwidth}{l}
 	\hline
 \end{tabular*}

Algorithm 3 converges to a local optimum after multiple iterations, as the updates in each iteration step of the algorithm are the optimal solutions to the corresponding subproblems. Then, Algorithm 3 converges to
\begin{equation}
	\begin{aligned}
		&\text{AR}_1\left(\mathbf{\Theta}^{(t)}, \mathbf{v}_1^{(t)}, \gamma^{(t)}, \mu^{(t)}\right)\\
		&\stackrel{(d)}{\le} \text{AR}_1\left(\mathbf{\Theta}^{(t)}, \mathbf{v}_1^{(t)}, \gamma^{(t)}, \mu^{(t+1)}\right)\\
		&\stackrel{(e)}{\le} \text{AR}_1\left(\mathbf{\Theta}^{(t)}, \mathbf{v}_1^{(t)}, \gamma^{(t+1)}, \mu^{(t+1)}\right)\\
		&\stackrel{(f)}{\le} \text{AR}_1\left(\mathbf{\Theta}^{(t)}, \mathbf{v}_1^{(t+1)}, \gamma^{(t+1)}, \mu^{(t+1)}\right)\\
		&\stackrel{(g)}{\le} \text{AR}_1\left(\mathbf{\Theta}^{(t+1)}, \mathbf{v}_1^{(t+1)}, \gamma^{(t+1)}, \mu^{(t+1)}\right)
	\end{aligned}
\end{equation}

where $(d)$, $(e)$, $(f)$, and $(g)$ are due to the update in $(\ref{11})$, $(\ref{12})$, $(\ref{13})$, and $(\ref{14})$, respectively. Moreover, $\text{AR}_1\left(\mathbf{\Theta}^{(t)}, \mathbf{v}_1^{(t)}, \gamma^{(t)}, \mu^{(t)}\right)$ has a finite upper bound since the limited power constraint. Therefore, we can guarantee the convergence of proposed Max-AR-CFFP algorithm.

The computational complexity of \textbf{Algorithm 3} is mainly determined by the updates of the four variables $\mu$, $\gamma$, $\mathbf{v}_1$, and $\mathbf{\Theta}$ via $(\ref{11})$, $(\ref{12})$, $(\ref{13})$, and $(\ref{14})$, respectively. The computational complexity of updating $\mu$ is $\mathcal{O}\left\{N\right\}$ FLOPs. The computational complexity of updating $\gamma$ is $\mathcal{O}\left\{M\right\}$ FLOPs. The complexity of updating $\mathbf{v}_1$ is $\mathcal{O}\left\{{\rm log_2(1/\delta)}\sqrt{M}(2M^4+M^3)\right\}$ FLOPs. The complexity of updating $\mathbf{\Theta}$ is $\mathcal{O}\left\{{\rm log_2(1/\delta)} \sqrt{N}(2N^4+N^3)\right\}$ FLOPs. Thus, the overall computational complexity of \textbf{Algorithm 3} is $\mathcal{O}\left\{L_c{\rm log_2(1/\delta)}(M^{4.5}+N^{4.5})\right\}$, wherein $\delta$ is the given accuracy tolerance of \textbf{Algorithm 3}, $L_c$ denotes the number of iterations required by \textbf{Algorithm 3} for convergence.

\rengSection{Simulation and Discussion}

In this section, simulation results are presented to prove the performance of the proposed two alternating iteration methods. Unless otherwise specified in the discussion, the parameters are set as follows. The locations of BS, active IRS, and user are set to (0 m, 30 m, 0 m), (50 m, 0 m, 10 m), and (25 m, 30 m, 0 m), respectively. Following \cite{g11}, the randomly generated channel matrix $\mathbf{G}$, channel vectors $\mathbf{f}$, and $\mathbf{h}$ follow the Rayleigh distribution. The number of BS antennas is chosen as follows:  $M=2$, noise power $\sigma_I^2 = \sigma^2_n = -100$ dBm.

\begin{figure}
	\begin{center}
		\includegraphics[width=1.0\columnwidth,keepaspectratio]{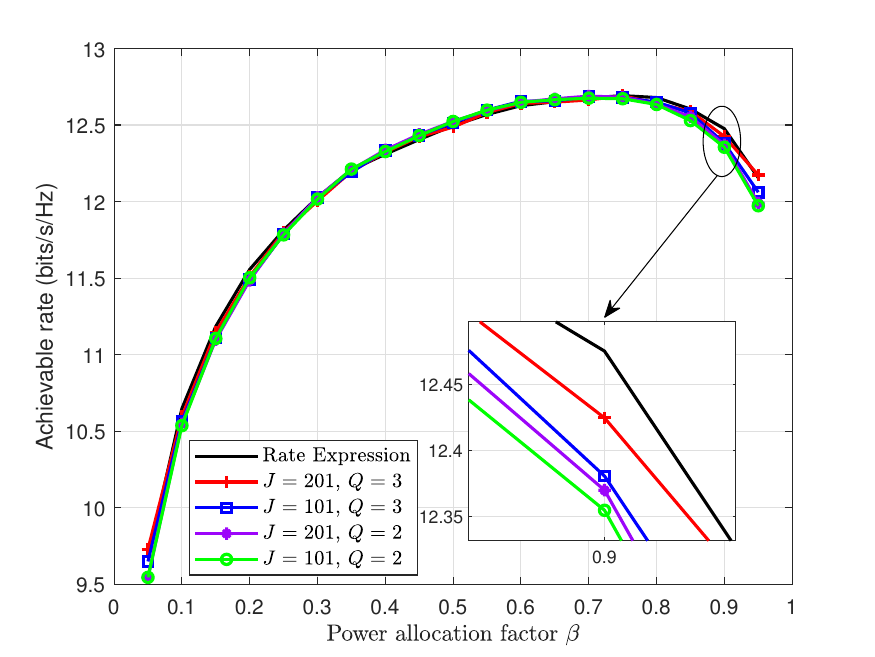}
		\fcaption{Achievable rate versus PA factor $\beta$.} 
	\end{center}
\end{figure}

Fig. 2 illustrates the curves of rate expression and its polynomial regression rate expression versus the PA factor $\beta$ for four distinct cases: $J=201, Q=3$; $J=101, Q=3$; $J=201, Q=2$; and $J=101, Q=2$ with $N=128$ and $P_{max}=30$ dBm. Channel path fading factors from BS to IRS, IRS to user, and BS to user are set to $\alpha_{BI}=2.1$, $\alpha_{IU}=2.1$, $\alpha_{BU}=4.0$, respectively. From Fig. 2, it can be seen that all four cases can fit the original rate curve well. As the number of sampling points $J$ and fitting order $Q$ increase, the polynomial regression fitting improves.

\begin{figure}
	\begin{center}
		\includegraphics[width=1.0\columnwidth,keepaspectratio]{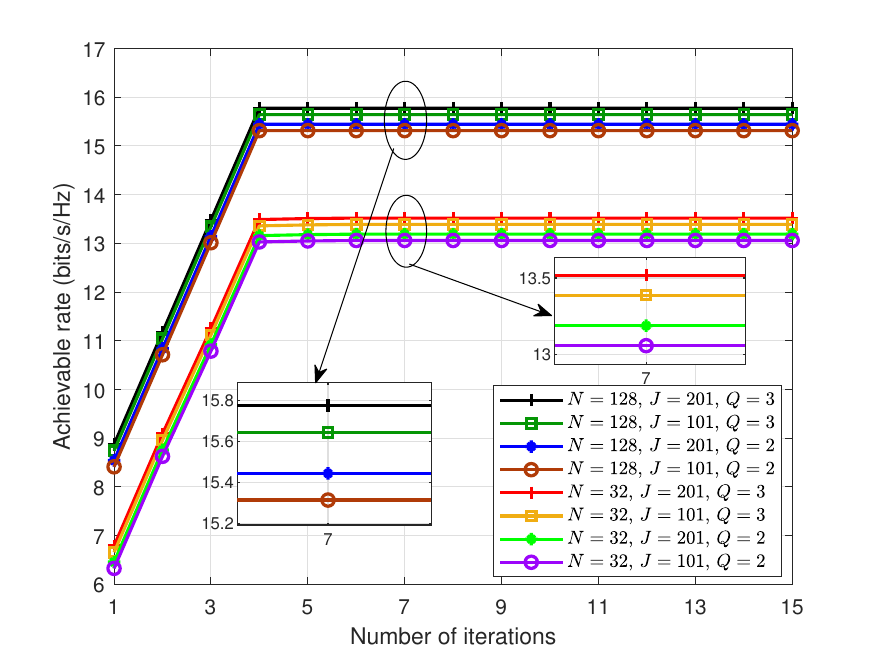}
		\fcaption{Convergence behaviour of the proposed Max-SNR-PA method.} 
	\end{center}
\end{figure}

Fig. 3 illustrates the convergence behaviour of the proposed Max-AR-PA method for two distinct active IRS phase shift elements: $N=32$ and $N=128$ with $P_{max}=30$ dBm, $\alpha_{BI}=2.1$, $\alpha_{IU}=2.1$, $\alpha_{BU}=4.0$. From Fig. 3, it is seen that the AR of proposed method increase rapidly with the number of iterations and finally converge to a value after a finite number of iterations. As $J$ and $Q$ increase, the AR performance of the proposed methods can be gradually improved. Considering that the rate difference between two cases: $J=201, Q=3$ and $J=101, Q=3$ is less than 0.2 bit, the number of sampling points $J$ and fitting order $Q$ are chosen to be $201$ and $3$.

\begin{figure}
	\begin{center}
		\includegraphics[width=1.0\columnwidth,keepaspectratio]{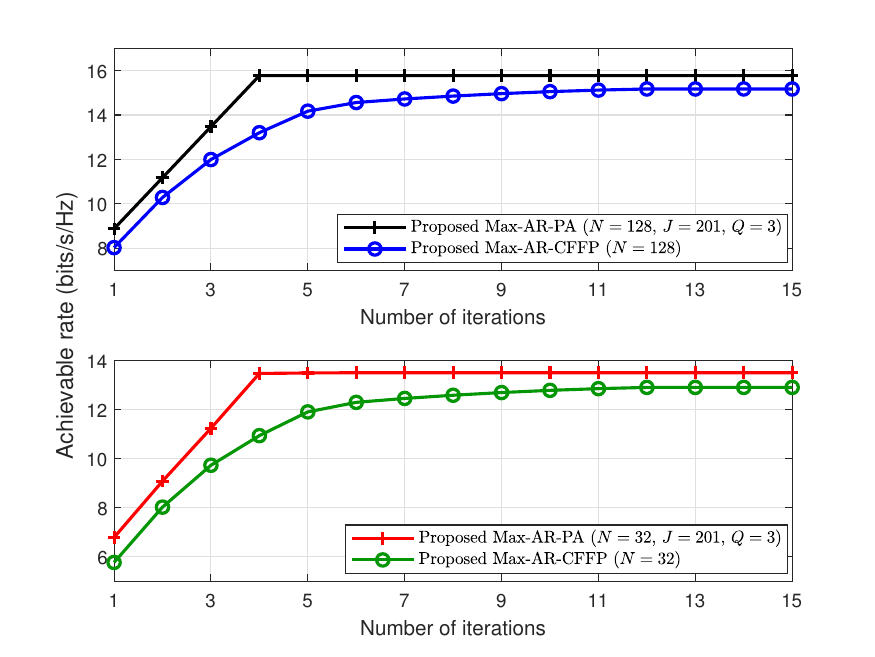}
		\fcaption{Convergence behaviour of the proposed two method.} 
	\end{center}
\end{figure}

Fig. 4 shows the convergence behaviour of the two proposed methods for two different active IRS phase-shifting elements: $N=32$ and $N=128$. Here, $P_{max}=30$ dBm, $\alpha_{BI}=2.1$, $\alpha_{IU}=2.1$, $\alpha_{BU}=4.0$. From Fig. 4, it is shown that as the number of iterations increases, the proposed Max-AR-CFFP method can also achieve convergence. But the convergence speed of Max-AR-CFFP method is slower than that of Max-SNR-PA method. 

\begin{figure}
	\begin{center}
		\includegraphics[width=1.0\columnwidth,keepaspectratio]{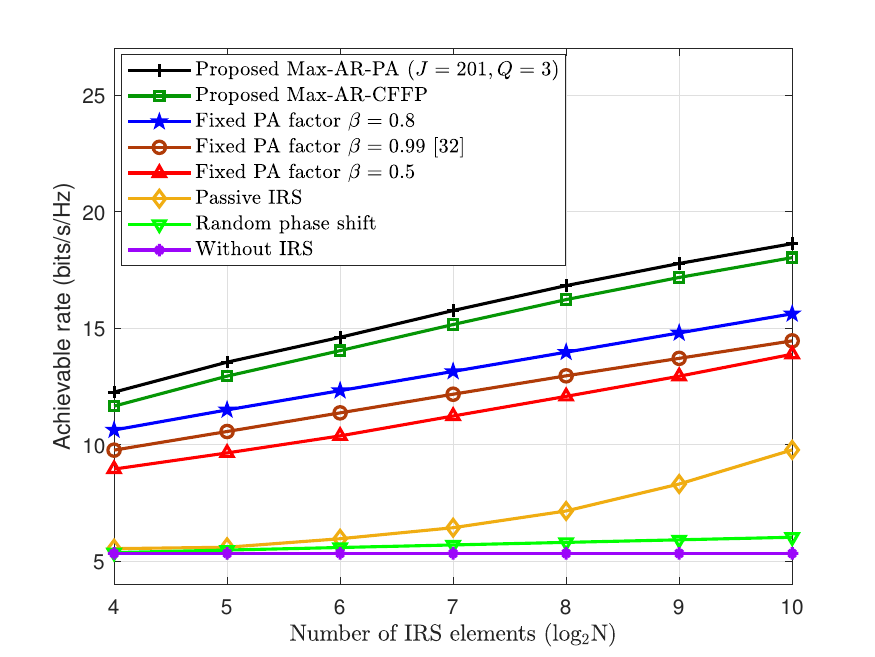}
		\fcaption{Achievable rate versus the number of active IRS elements $N$ with a weak direct link.} 
	\end{center}
\end{figure}

\begin{figure}
	\begin{center}
		\includegraphics[width=1.0\columnwidth,keepaspectratio]{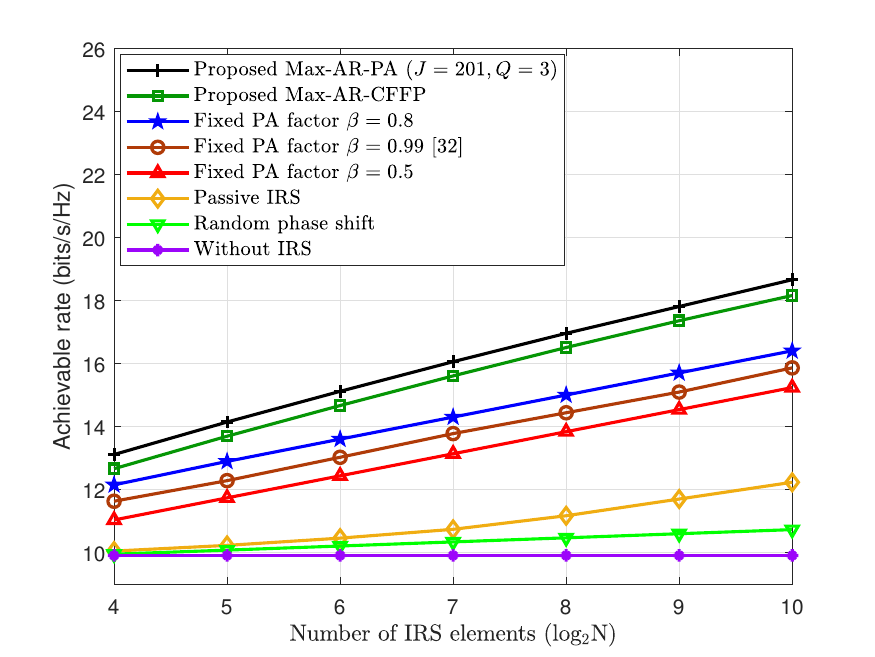}
		\fcaption{Achievable rate versus the number of active IRS elements $N$ with a medium strength direct link.} 
	\end{center}
\end{figure}

\begin{figure}
	\begin{center}
		\includegraphics[width=1.0\columnwidth,keepaspectratio]{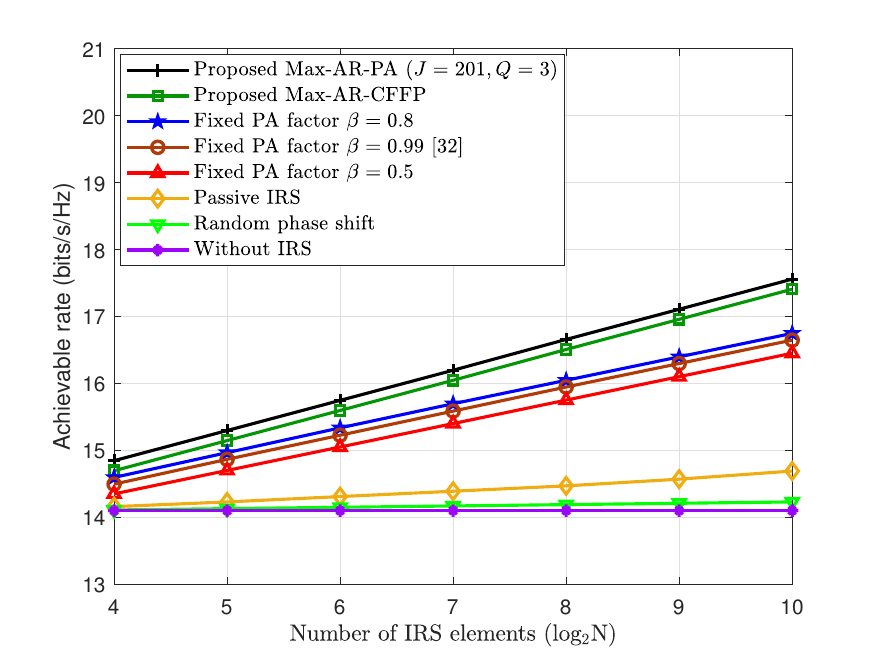}
		\fcaption{Achievable rate versus the number of active IRS elements $N$ with a strong direct link.} 
	\end{center}
\end{figure}

Fig. 5, Fig. 6, and Fig. 7 plot the achievable rates of the proposed Max-SNR-PA and Max-AR-CFFP methods versus the number of active IRS elements $N$ with fixed PA factor $\beta=0.8$,  fixed PA factor $\beta=0.99$ \cite{g32}, fixed PA factor $\beta=0.5$, passive IRS, random phase shift, and without IRS as performance benchmarks. Here, $P_{max} = 30$ dBm, $\alpha_{BI}=2.1$, $\alpha_{IU}=2.1$. The channel path fading factor from BS to user in Fig. 5, Fig. 6, and Fig. 7 are set to $\alpha_{BU}=4.0$, $\alpha_{BU}=3.0$, and $\alpha_{BU}=2.1$, respectively. This means that the direct link strength from BS to user are weak, medium and strong, respectively. 

From these three figures, we can see that as the number of IRS elements $N$ increases, the rates of all eight methods will be promoted. The proposed two methods can achieve an obvious rate performance gains over fixed PA factor $\beta=0.8, \beta=0.99, \beta=0.5$, passive IRS, random phase shift, and without IRS. Especially, with the enhancement of direct link between BS and user, the improvement effect of passive IRS on rate performance gradually deteriorates, and the two proposed PA strategies can still effectively improve rate performance.

\begin{figure}
	\begin{center}
		\includegraphics[width=1.0\columnwidth,keepaspectratio]{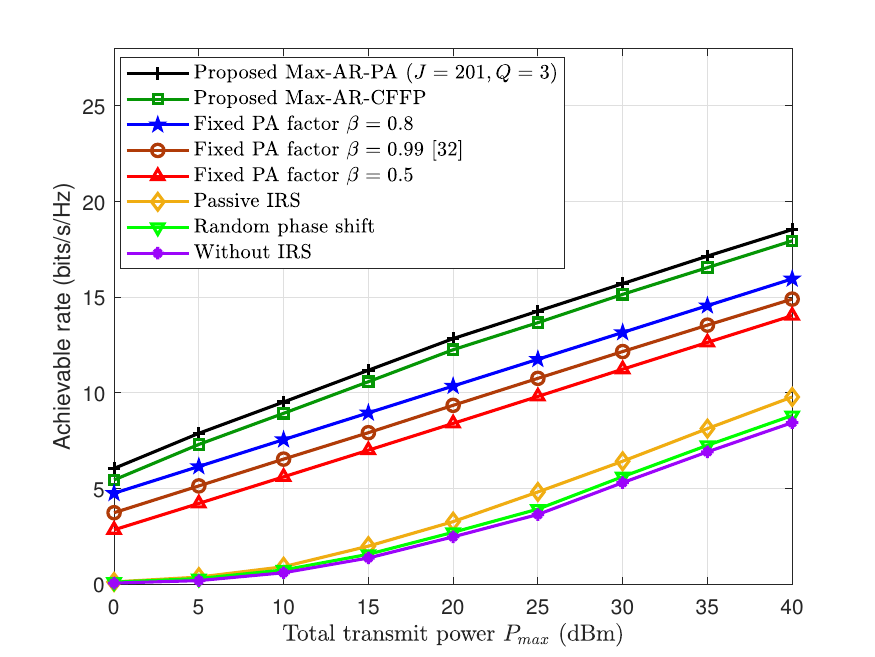}
		\fcaption{Achievable rate versus total transmit power $P_{max}$ with a weak direct link.} 
	\end{center}
\end{figure}

\begin{figure}
	\begin{center}
		\includegraphics[width=1.0\columnwidth,keepaspectratio]{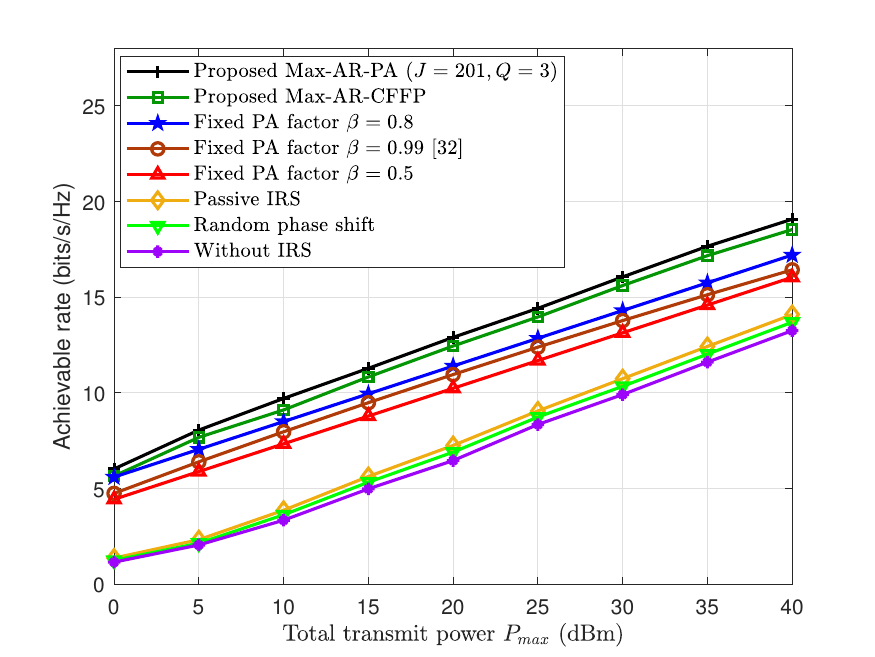}
		\fcaption{Achievable rate versus total transmit power $P_{max}$ with a medium strength direct link.} 
	\end{center}
\end{figure}

\begin{figure}
	\begin{center}
		\includegraphics[width=1.0\columnwidth,keepaspectratio]{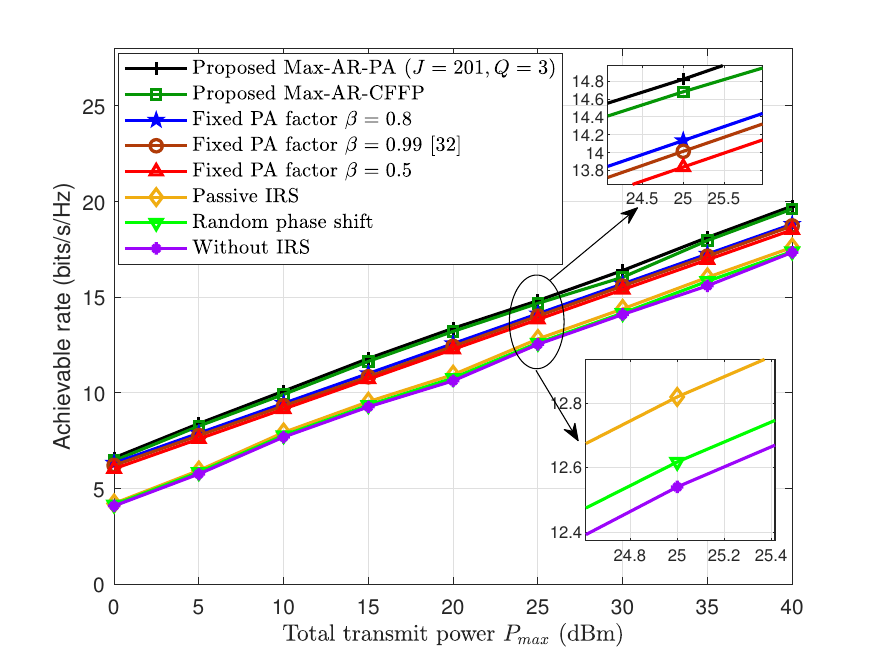}
		\fcaption{Achievable rate versus total transmit power $P_{max}$ with a strong direct link.} 
	\end{center}
\end{figure}

Fig. 8, Fig. 9, and Fig. 10 illustrate the curves of achievable rate versus total transmit power $P_{max}$, where $N=128$, $\alpha_{BI}=2.1$, $\alpha_{IU}=2.1$. Here, $\alpha_{BU}$ are set to $4.0$, $3.0$, and $2.1$ in Fig. 8, Fig. 9, and Fig. 10, respectively. These three figures is shown that as $P_{max}$ increases, all eight curves in figures has an upward trend. When the direct link between BS and user is gradually enhanced, the proportion of signal strength transmitted through direct link in user received signals will be gradually improved. Thus, the rate performance advantages of two proposed PA strategies are more prominent in the cases of weak direct link and medium strength direct link. In summary, eight methods have an increasing order in rate performance as follows: Max-AR-PA, Max-AR-CFFP, fixed PA factor $\beta=0.8$, fixed PA factor $\beta=0.99$ \cite{g32}, fixed PA factor $\beta=0.5$, passive IRS, random phase shift, and without IRS.

\rengSection{Conclusion}

In this paper, we investigated the AR performance of an active IRS-aided wireless network with PA. To improve the AR performance, under a limited total power constraint, a SNR maximization problem was constructed by jointly optimizing PA factor, active IRS phase shift matrix, and BS beamforming vector. To address the non-convex problem, an alternating optimization method was used. Specifically, the PA factor was obtained by polynomial regression method, BS and IRS beamformings were derived based on Dinkelbach's transform and successive convex approximation techniques. To reduce the computational complexity of above proposed strategy, we maximize AR and alternately optimize BS and IRS beamformings by a closed-form fractional programming method. Simulation results proved that our proposed two strategies obviously outperform the benchmark schemes and can achieve significant AR performance gains.

\rengAck

This work was supported in part by the National Natural Science Foundation of China (Nos.U22A2002, and 62071234), the Hainan Province Science and Technology Special Fund (ZDKJ2021022), the Scientific Research Fund Project of Hainan University under Grant KYQD(ZR)-21008, the Collaborative Innovation Center of Information Technology, Hainan University (XTCX2022XXC07), and the National Key Research and Development Program of China under Grant 2023YFF0612900.

\vspace{0cm}
\begin{center}
\noindent{\Large\bfseries About the Authors \dots}
\vspace{0cm}
\end{center}

\noindent\textbf{Qiankun CHENG} received the B.E. degree from Anhui Normal University, China, in 2022. He is currently pursuing the M.S. degree with the School of Information and Communication Engineering, Hainan University, China. His research interests include intelligent reflecting surface and machine learning.

\noindent\textbf{Jiatong BAI} (corresponding author) is currently pursuing the Ph.D. degree with the School of Information and Communication Engineering, Hainan University, China. Her research interests include wireless communication and machine learning.

\noindent\textbf{Baihua SHI} (corresponding author) is currently pursuing the Ph.D. degree with the School of Electronic and Optical Engineering, Nanjing University of Science and Technology, China. His research interests include wireless communication and signal processing.

\noindent\textbf{Wei GAO} received the B.E. degree in communication engineering and the Ph.D. degree of information and communication engineering from the Huazhong University of Science and Technology (HUST) in 2014 and 2020, respectively. He is
currently a postdoc with Hainan University. His research interests include network architecture, wireless network access and radio resources allocation.

\noindent\textbf{Feng SHU} (Member, IEEE) was born in 1973. He received the B.S. degree from Fuyang Teaching College, Fuyang, China, in 1994, the M.S. degree from Xidian University, Xian, China, in 1997, and the Ph.D. degree from Southeast University, Nanjing, China, in 2002. From 2009 to 2010, he was a Visiting Postdoctoral Fellow with the University of Texas at Dallas, Richardson, TX, USA. From July 2007 to September 2007, he was a Visiting Scholar with the Royal Melbourne Institute of Technology, Melbourne VIC, Australia. From 2005 to 2020, he was with the School of Electronic and Optical Engineering, Nanjing University of Science and Technology, Nanjing, where he was promoted from an Associate Professor to a Full Professor of supervising Ph.D. students in 2013. Since 2020, he has been with the School of Information and Communication Engineering, Hainan University, Haikou, China, where he is currently a Professor and a Supervisor of Ph.D. and graduate students. He has authored or coauthored more than 300 in archival journals with more than 150 papers on IEEE journals and 250 SCI-indexed papers. His research interests include wireless networks, wireless location, and array signal processing. He was awarded with the Leading-Talent Plan of Hainan Province in 2020, the Fujian Hundred-Talent Plan of Fujian Province in 2018, and the Mingjian Scholar Chair Professor in 2015. His citations are more than 8000 times. He holds one US patent and more than 40 Chinese patents. He is also a PI or CoPI for eight national projects. He was an Exemplary Reviewer for IEEE Transactions on Communications in 2020. He is currently the Editor of IEEE Wireless Communications Letters and guest editors for the journals Chinese Journal of Aeronautics and Journal of Electronics Information Technology etc. He was the Editors of IEEE Systems Journal from 2019 to 2021 and IEEE Access from 2016 to 2018 and also guest editors for IET Communications and Security and Safety etc.
\end{multicols}
\end{document}